\documentclass{elsart}

\usepackage{graphicx}
\usepackage{epsfig}
\usepackage{amssymb}
\usepackage{color}

\newcommand{\eqnref}[1]{Eq.~(\ref{#1})}

\newcommand{\be}{\begin{eqnarray}}
\newcommand{\ee}{\end{eqnarray}}

\newcommand{\nn}{\nonumber }

\newcommand{\beq}{\begin{equation}}
\newcommand{\eeq}{\end{equation}}
\newcommand{\bea}{\begin{eqnarray}}
\newcommand{\beas}{\begin{eqnarray*}}
\newcommand{\beau}[1]{\begin{equation} \label{#1} \begin{array}{rcl}}
\newcommand{\eea}{\end{eqnarray}}
\newcommand{\eeas}{\end{eqnarray*}}
\newcommand{\eeau}{\end{array} \end{equation}}
\newcommand{\bay}{\begin{array}}
\newcommand{\eay}{\end{array}}
\newcommand{\bals}{\begin{align*}}
\newcommand{\eals}{\end{align*}}

\begin{document}

\begin{frontmatter}



  \title{Transverse Momentum Broadening \\  in Semi-inclusive DIS on  Nuclei}

   \author[Heidelberg]{S.~Domdey\thanksref{domdeymail}},
  \author[Heidelberg]{D.~Gr\"unewald\thanksref{Grunemail}},
   \author[Valparaiso,Heidelberg]{B.Z.~Kopeliovich\thanksref{Kopeliovichmail}}
   and \author[Heidelberg]{H.J.~Pirner\thanksref{Piremail}}


  \address[Heidelberg]{Institut f\"ur Theoretische Physik, Philosophenweg 19, D-69120 Heidelberg, Germany }
  \address[Valparaiso]{Departamento de F\'\i sica y Centro de Estudios
Subat\'omicos,\\ Universidad T\'ecnica
Federico Santa Mar\'\i a, Casilla 110-V, Valpara\'\i so, Chile}

  \thanks[domdeymail]{E-mail address: domdey@tphys.uni-heidelberg.de }
  \thanks[Grunemail]{E-mail address: daniel@tphys.uni-heidelberg.de}
  \thanks[Kopeliovichmail] {E-mail address: boris.kopeliovich@usm.cl}
  \thanks[Piremail]{E-mail address: pir@tphys.uni-heidelberg.de}

  \begin{abstract}
Using a three  stage model of hadron formation we calculate the
change of the transverse momentum distribution of hadrons produced
in semi-inclusive deep inelastic scattering (SIDIS)  on nuclei. In
the first stage after its interaction with the virtual photon, the
struck quark propagates quasi free in the nuclear environment
undergoing multiple collisions with nucleons. During this stage it
can acquire transverse momentum. In the second stage a prehadron
is formed which has a very small elastic cross section with the
nucleons. In the third stage the prehadron turns into a hadron.
For HERMES energies, prehadron elastic scatterings contribute
little to $p_\perp$-broadening. The acquired extra $\Delta
p_{\bot}^2$ of hadrons can therefore be deduced entirely from the
first stage of quasi free quark propagation with quark-nucleon
collisions. We use this model to describe $\pi$-production on Ne,
Kr, Xe and compare with the most recent HERMES preliminary data.

  \end{abstract}



\end{frontmatter}


\baselineskip18pt
\section{Introduction}

Fragmentation of quarks and gluons into hadrons is a consequence of color confinement and is therefore one of the most-interesting parts of non-perturbative QCD.
Use of nuclear targets allows the experimentalist to position detectors  (nucleon targets) of the hadro\-ni\-za\-tion process near the interaction point and probe hadron formation on length scales of a few Fermi.

A struck quark originated from deep-inelastic scattering (DIS) propagates through the nucleus experiencing multiple interactions which cause induced energy loss. In the string model the medium-induced energy loss results from production of new strings in multiple collisions of the quark \cite{bk90}. In a perturbative description the excess of energy loss is related to medium-induced gluon radiation resulting from broadening of the quark transverse momentum \cite{baier}. In both approaches induced energy loss rises quadratically with the path-length of the quark.

Although induced energy loss contributes to the attenuation of the produced hadron, even  stronger suppression may result from absorption of the produced hadron, if the struck quark is neutralized and a  colorless dipole is formed inside the nucleus.
In the string model, the production of a hadron may be described as a two stage process \cite{Accardi:2002tv,Falter:2004uc,Accardi:2005jd}.
First a prehadron (e.g. a dipole of small size) is created, which has a reduced absorption cross section, and later the hadron wave function is formed. These models \cite{Accardi:2002tv,Falter:2004uc,Accardi:2005jd} put the main emphasis on the effects of absorption and neglect the induced energy loss.
The attenuation of the produced hadron can be described more consistently within the color-dipole approach employing a path integral technique \cite{boris,Kopeliovich:2004kq} which describes the evolution of the dipole propagating through the medium. The induced energy loss evaluated perturbatively in Ref.~\cite{baier} was included in Ref.~\cite{Kopeliovich:2004kq}.

Some models put the main emphasis on the effects of induced energy loss, assuming that color neutralization of the quark always occurs at long distance from the DIS location, i.e. outside of the nucleus \cite{Wang:2002ri,Arleo:2005wq}. Of course, this assumption may only be valid in certain kinematical domains.

A novel mechanism of hadron attenuation was found recently in Ref.~\cite{Kopeliovich:2008uy}.
It turns out that even if no induced energy loss occurs, and color neutralization happens outside of the nucleus, a significant nuclear suppression can result from quantum-mechanical interferences between different amplitudes.

In this paper we
study the
recent HERMES preliminary results \cite{VanHaarlem:2007kj} which
give preliminary data on $p_{\bot}$-broadening of pions produced in deep inelastic
lepton-nucleus scattering (DIS) on Ne, Kr and Xe nuclei.
Transverse broadening $\Delta p_{\bot}^2=\langle p_\perp^2\rangle_A-\langle p_\perp^2 \rangle_D$ 
in the nuclear medium of a nucleus with mass number $A$ has been measured as a function
of the hadron fractional momentum $z_h$,
of the virtual photon energy $\nu$ and its virtuality $Q^2$, for different
nuclear size and for different hadrons. Here, the index $D$ in the definition of the transverse broadening refers to a deuteron target.

We use the absorption model \cite{Accardi:2002tv,Accardi:2005jd} to calculate the nuclear modifications of hadron
production in DIS.
The hadron formation time is computed analytically
in the framework of the LUND string fragmentation model as a three-step process. In
the first stage the quark (or antiquark) ejected from the nucleon propagates and undergoes
multiple collisions in the nucleus. In the second stage color neutralization takes place and a
prehadron is formed. Inelastic interactions of the prehadron or hadron result in a considerable
shift of the final (detected) hadron towards smaller $z_h$. We treat this process as absorption.
Only elastic rescatterings preserve the (pre-)hadron and contribute to broadening. However, the
elastic cross sections of hadrons and prehadrons are very small compared to the inelastic cross
sections.
In the third stage the final state hadron is formed from the surviving prehadrons.
With an inelastic prehadron nucleon cross section which is reduced compared to the
hadron-nucleon cross section, the model showed  rather good agreement with the available HERMES
data \cite{HERM01,HERM03,HERM04} for pions and kaons.
In the LUND model, the hadron is formed at the formation length $l_h$
\begin{equation}
 l_h=l_p+z_h \frac{\nu}{\kappa}\,.
\end{equation}
Here, $l_p$ denotes the prehadron formation or production length.
For relativistic quarks confined in one dimension the only scale
setting parameter is the string tension $\kappa$. Therefore, in the
conventional estimate of the Lund model for the formation length
enters the photon energy divided by the string tension
\begin{equation}
L=\frac{\nu}{\kappa}~~,~~\kappa=1 \frac{\mbox{GeV}}{\mbox{fm}}\,.
\end{equation}

We have fitted the prehadron cross
section to the pion-data for the multiplicity ratios as a function of $z_h$ and $\nu$
and found an optimal fit with a prehadron cross section equal to (2/3)
of the hadron cross section in the extended modelling of Ref.~\cite{Accardi:2005jd}
\begin{equation}
\sigma_{prehadron}\approx \frac{2}{3}\, \sigma_{hadron}.
\end{equation}
This value of the prehadronic cross section is in agreement
 with Ref.~\cite{Falter:2004uc}.
Such a reduction may be partially because the prehadron does not
yet have the full size of the hadron and therefore interacts with
a smaller cross section due to color transparency. Besides, a
considerable reduction should be also expected,  since the
exponential attenuation of (pre)hadrons used in the fit (see below
Eq.~(\ref{1})) misses the Gribov inelastic corrections
\cite{gribov}, which make nuclei much more transparent
\cite{zkl,Kopeliovich:1995yx,transparent}.

Further on, we have considered in \cite{Accardi:2005jd} 
the production process for particles which cannot be formed from
a valence quark in the proton which is knocked out by the photon.
For example,
negative kaons as well as antiprotons cannot be
formed by a struck valence quark picking up an antiquark
from the string break-up. They can only be formed from struck sea quarks,
which are subdominant at HERMES, or from $q\bar q$
pairs formed inside the colour string.
This different mechanism implies a
flavour-dependent formation time from the LUND string fragmentation model.

\section{Hadronic $p_\perp$-broadening as a function of $z_h$ and $\nu$}

In this paper we want to test the three stage model further by
concentrating on quark propagation through the nucleus  where most of the transverse momentum
is acquired.
The theoretical calculation is very similar to reference \cite{Accardi:2002tv}.
The length $l_p$ after which the prehadron is formed
 \cite{Accardi:2002tv,Accardi:2005jd} depends on the
energy $\nu$ transferred to the quark, the string tension $\kappa$
and the energy fraction $z_h$ of the produced hadron. From energy
conservation already follows that if the hadron has a very large
$z_h$ the quark cannot have radiated very much energy. Therefore
the formation length of the colour neutral state must have been
very short \cite{bk90,kn,bg}. The prehadron formation length is
computed analytically in the framework of the LUND string
fragmentation model. If one assumes that the prehadrons can be
formed directly from the struck quark by picking up an antiquark
from the string break up, then the prehadron formation length
reads
\begin{eqnarray}\label{lp}
l_p&=&\frac\nu\kappa z_h (1 -
   z_h) \nonumber\\
   & &
   \times \displaystyle\left[
    1 +
    \frac{1 + D_q}{2 + D_q}\frac{ 1 - z_h}{z_h^{2 + D_q}}
       \mbox{ }_2F_1\left(2 + D_q, 2 + D_q, 3 + D_q, \frac{z_h-1}{z_h}\right)\right]\,.
\end{eqnarray}

Here the parameter $D_q$ is equal to $D_q=0.3$ and ${}_2F_1$ is the Gauss hypergeometric function.
The corrections to the simple $z_h(1-z_h)$-behavior of the prehadronic formation length
$l_p$ given by the Gauss hypergeometric function can be recasted into effective powers of $z_h$ and  $1-z_h$ normalized by an appropriate prefactor. One can obtain
an excellent fit to the "exact" prehadron formation length by
\begin{equation}
l_p \simeq 1.19\,\, \frac\nu\kappa\, z_h^{0.61}\, (1 -
   z_h)^{1.09}.
\end{equation}
We use the length of the quark trajectory \eqnref{lp} to calculate with the dipole model the
acquired $\Delta p_{\bot}^2$ of the quark under the constraint that the subsequent prehadron
is not absorbed on its a way through the nucleus, i.e. that it can be finally detected as a hadron.
This is necessary, since the information about the acquired transversal momentum of the struck quark is encoded in the detected hadrons only.
\begin{eqnarray}\label{deltaptq}
\langle\Delta p_\perp^2\rangle_q &=& \langle\sigma q_{\bot}^2\rangle \frac{1}{\langle S_*\rangle}
\int_{-\infty}^{\infty} d^2b\,dz\, \rho_A\left(\vec{b},z\right)
\int_z^{z+l_p} dz' \rho_A\left(\vec{b}, z'\right)\nn\\
& &~~~~~~~~~~~~~\cdot\exp\left(-\sigma_*\,\int_{z+lp}^\infty
dz^{\prime\prime} \,\rho_A\left(\vec{b},
z^{\prime\prime}\right)\right)\,,\nonumber\\ \langle S_*\rangle
&=& \int_{-\infty}^{\infty} d^2b\,dz\,
\rho_A\left(\vec{b},z\right)\,\exp\left(-\sigma_*\,
\int_{z+lp}^\infty dz^{\prime} \,\rho_A\left(\vec{b},
z^{\prime}\right)\right)\,. \label{1}
\end{eqnarray}
%
%
%
In this equation, the quantity $\langle\sigma q_{\bot}^2\rangle$
is the mean transverse momentum squared $q_{\bot}^2$ acquired by
the quark in one collision multiplied with the corresponding quark
nucleon cross section. This quantity is related to the dipole
nucleon cross-section \cite{Johnson:2000dm,Kopeliovich:2000ra} as
shown below. Furthermore, we assume a sharp distribution of
prehadron formation points, namely the prehadron is produced after
travelling a distance $l_p$ through the nucleus. Hence, the final
induced momentum broadening calculated in \eqnref{1} can be read
as the mean transverse momentum $q_{\bot}^2$ acquired by the
ejected quark in a single collision multiplied with the average
number of collisions in the nucleus. The resulting transverse
momentum is averaged over all virtual photon interaction points
and weighted by the prehadron survival probability. To be precise,
the first integral over the nuclear density $\rho_A$ averages over
all primary interaction points in which a quark is ejected from a
nucleon. The second integral multiplied with the cross section
$\sigma$ yields the number of collisions suffered by the ejected
quark. The exponential factor at the end represents the prehadron
survival probability $S_*$ \cite{Accardi:2005jd}. It is  dictated
by the longitudinal thickness of the nucleus at a given impact
parameter $b$ and by the mean free path of the prehadron in the
nucleus $\lambda_*^{-1}=\sigma_*\,\rho_A$, where the prehadron
nucleon cross-section $\sigma_*=2/3\, \sigma_{\pi N}$ (see
Ref.~\cite{Accardi:2005jd}). The mean free path $\lambda_*$ is of
the same magnitude as $l_p$ and $R_A$. Hence, more weight is given
to production points close to the back-surface of the nucleus
which have large prehadron survival probabilities. In order to
normalize this expression to the actual number of detected hadrons, we
divide by the $z$-integrated prehadron survival factor $\langle
S_*\rangle$.

The mean transverse momentum squared times the cross-section,
i.e. $\langle\sigma q_{\bot}^2\rangle$, can be derived from the dipole
nucleon cross section as follows.  In the eikonal approximation, the
ejected high momentum parton moves on a classical trajectory with
impact parameter $\vec{b}$ and picks up a non-abelian phase factor
$V(\vec{b})$ in the background gauge field generated by the nucleon
\begin{eqnarray}
V(\vec{b})&=&{\cal P}{\,}\exp\left[\mathrm{i}\,g\int_{-\infty}^{+\infty}dx^\mu\, A_\mu(x)\right]\,.
\end{eqnarray}
Here $V(\vec{b})$
is the Wilson line of the parton with impact parameter $\vec{ b}$ relative to the proton.
We use the notation $A_\mu\equiv A_\mu^a\,t^a$, where
$t^a$'s are the generators of the group SU($N_c$) in the fundamental representation.
The differential cross section to produce a parton with transverse momentum $\vec{q}_\perp$ is
given by projecting the eikonal phase onto $\vec{q}_\perp$ and by
taking the modulus of the amplitude integrated over all possible impact parameters
\begin{equation}
\frac{d\sigma}{d^2q_\perp}=\frac{1}{(2\,\pi)^2}\int d^2b\,d^2b^\prime
{\rm e}^{i\,\vec{ q}_\perp(\vec{b}-\vec{b}^\prime)}{\,}\frac{1}{N_c}\left\langle{\rm Tr}\left[V^\dagger(\vec{b}^\prime)\,V(\vec{b}) \right]\right\rangle\,.
\label{eqn:PtCrossSec}
\end{equation}
Hence, a fake dipole of size
$\vec{r}_{\bot}=\vec{b}-\vec{b}^\prime$ is constructed from the
ejected parton in the $V$-amplitude and in the
$V^\dagger$-amplitude. Their trajectories  are displaced from each
other by the distance $r_{\bot}$. The expectation values of the
Wilson lines have to be evaluated with respect to the target
ground state.
In the dipole model, the total cross-section for the interaction
of a dipole of size $\vec{r}_{\bot}$ with a target nucleon is given by
\begin{eqnarray}
\sigma_{dN}(\vec{r}_{\bot})&=&2\,
\int d^2b \left(1-\frac{1}{N_c}\left<{\rm Tr}\left[V^\dagger(\vec{b}+\vec{r}_\perp)\,V(\vec{b})\right]\right>\right)\,.
\end{eqnarray}
We define the quantity $\left<\sigma q_\perp^2\right>$ as the integral over transverse momentum $d^2q_\perp$ of the differential cross section given in \eqnref{eqn:PtCrossSec} multiplied by $q_\perp^2$.
Differentiating
the phase factor appearing in \eqnref{eqn:PtCrossSec} twice
with respect to the transversal separation and performing the integral over $d^2q_\perp$ one sees that
$\left<\sigma q_\perp^2\right>$ is related to the dipole nucleon cross section.
\begin{eqnarray}
\left<\sigma q_\perp^2\right>
&\equiv& \int d^2q_\perp \frac{d\sigma}{d^2q_\perp}q_{\perp}^2\nn\\
&=&\frac{1}{(2\,\pi)^2}\int d^2q_\perp \int d^2b\,d^2r_\perp \left(-\nabla_{\bot}^2
{\rm e}^{i\vec{ q}_\perp\vec{ r}_{\bot}}\right){\,}\frac{1}{N_c}\left\langle{\rm Tr}\left[V^\dagger(\vec{b}+\vec{r}_\perp)\,V(\vec{b}) \right]\right\rangle\nn\\
&=&\frac{1}{2}\left. \nabla_{\bot}^2 \sigma_{dN}(\vec{r}_{\bot})\right|_{r_\perp=0}.
\label{broad}
\end{eqnarray}
This expression confirms the result derived in \cite{Johnson:2000dm}.
Because of the $q_\perp$-integration only the second derivative of the $r_\perp^2=0$-part
in the dipole cross section is relevant for $p_\perp$-broadening.
This derivative is a constant due to the color transparency behavior of the dipole cross section $\sigma_{dN}(\vec{r}_{\bot})_{r_\perp\to0}\propto r_{\bot}^2$ \cite{zkl}.
We use a form
of the $x$-dependent phenomenological dipole nucleon cross section of
Ref.~\cite{GolecBiernat:1998js},
which has been adjusted to include soft interactions in Ref.~\cite{Kopeliovich:1999am}
\begin{equation}
\sigma_{dN}(\vec{r}_{\bot})=\sigma_0(s)\left[1-\exp\left(-\frac{\vec{r}_\perp{}^2}{r_0^2(s)} \right) \right]\,,
\label{kst}
\end{equation}
where $r_0(s)=0.88\,(s/s_0)^{-0.14}\,\mathrm{fm}$,
$s_0=1000\,\mathrm{GeV}^2$ and
\begin{equation}
\sigma_0(s)=23.6\,\left(\frac{s}{s_0}\right)^{0.08}\left(1+\frac38
\frac{r_0^2(s)}{0.44\,\mathrm{fm}^2}\right)\,\mathrm{mb}.
\end{equation}
Hence, one determines the parameter $\langle\sigma
q_{\bot}^2\rangle\simeq 4.6$ for $\sqrt{s}\simeq 5\,\mathrm{GeV}$,
which is the typical $\sqrt{s}$ for the quark nucleon scattering
at HERMES energies.

Notice that the quark-nucleon differential cross section $\sigma$ is infrared divergent. For this reason the mean momentum transfer squared $\langle q_T^2\rangle=0$. Nevertheless, the broadening,
$\langle\sigma q_T^2\rangle$, is nonzero and finite. It results from infinitely many soft rescatterings with vanishingly small momentum transfers. One can introduce an ad hoc infra-red cutoff and regularize the problem. However, this cutoff does not affect the final result Eq.~(\ref{broad})
\cite{Johnson:2000dm}.
The new scale controlling broadening is called saturation momentum $Q_s$. For a quark propagating a path-length $L$ in nuclear medium of density $\rho_0$ the saturation momentum reads
\begin{equation}
(Q_s^A)^2={1\over2}\,\rho_0\,L\,\sigma_0\,(Q_s^N)^2,
\label{Qs}
\end{equation}
where $\sigma_0$ and $(Q_s^N)^2=4/r_0^2$ are defined in terms of the saturated form \cite{GolecBiernat:1998js,Kopeliovich:1999am} of the dipole cross section, Eq.~(\ref{kst}).
This increase in saturation scale is identical to the broadening of the
mean transverse momentum squared  of the quark. One should mention
that $Q_s^A$ is the saturation scale for quarks. For gluons, the
saturation scale squared is $9/4$ times larger. A review of other
approaches to nuclear broadening can be found in
Ref.~\cite{Raufeisen:2003zk}. In this reference, the relevant
quantity is $\hat q_F=\langle\sigma q_{\bot}^2\rangle \rho_0$
which is given as $ \hat q_F=0.035 $ GeV$^2$/fm compared with our
determination  $\hat q_F=0.032 $ GeV$^2$/fm  at this dipole
energy.

The acquired transverse momentum of the quark given in \eqnref{1}
can be computed analytically if one uses a hard sphere
approximation for the target nucleus which has a homogeneous nuclear
density $\rho_0=\left(4\,\pi/3\,r_0^3\right)^{-1}$ with
$r_0=1.2$ fm
\begin{equation}
\rho_A(b,z)=\rho_0\,\Theta(R_A-b)\,\Theta(R(b)-|z|)~~,~~R(b)=\sqrt{R_A^2-b^2}\,.
\label{eqn:HSDens}
\end{equation}
Cylindrical coordinates $(b,z)$ are favorable due to rotational invariance in the impact parameter plane. Here, the impact
parameter of the initial virtual photon is $b$ and $R_A=r_0\,A^{1/3}$ denotes the nuclear radius.
The hard sphere approximation is reasonable for large nuclei in which the thickness
of the boundary of the nucleus is small in comparison to its extension.
This approximation gives us some insights into the underlying physics of the $p_\perp$
broadening
\newpage
\begin{eqnarray}
\lefteqn{ \langle\Delta p_\perp^2\rangle_q = \langle\sigma
q_{\bot}^2\rangle\,\rho_0 \Bigg\{ l_p\Bigg[1-\frac{1}{\langle
S_*\rangle}\cdot\Bigg(
\frac{3}{8}\frac{l_p}{R_A}-\frac{1}{64}\left(\frac{l_p}{R_A}\right)^3
\Bigg)\Bigg]\,
\Theta(2\,R_A-l_p)\Bigg.}&&\nonumber\\
& & ~~~~~~~~~~~~~~~~~~~~~~~~~~~~~~~~~~~~~~~~~~~~~~~~~~~~~
\Bigg.+\frac{3}{4}R_A\,\Theta(l_p-2\,R_A)
\Bigg\}\,.
\end{eqnarray}
The $p_\perp^2$-broadening is given by the acquired mean momentum squared
per unit length $\langle\sigma q_{\bot}^2\rangle\,\rho_0$ times the in-medium
propagation length of the quark multiplied with the normalized
survival probability of the prehadron.
This product is represented by the expression in square brackets.

The in-medium propagation length of the quark has two
contributions which differ depending on the relation between the
prehadron formation length $l_p$ and the nuclear diameter
$2\,R_A$. If the prehadron formation length is larger than the
nuclear diameter, then the prehadron is formed outside of the
nucleus and the in-medium propagation length of the quark is given
by the average nuclear thickness $3/4\,R_A$ seen by the quark. The
prehadron survival probability is equal to one in this case. For
prehadron formation lengths which are smaller than the nuclear
diameter, the in-medium propagation length of the quark equals the
prehadron formation length $l_p$ plus some higher order
corrections in $l_p/R_A$ which are due to the finite size of the
nucleus. These corrections account for the possibility that the
prehadron is formed outside of the nucleus such that not the
entire prehadron formation length $l_p$ contributes. The finite
size corrections are normalized by the prehadron survival
probability, whereas the expression which one would expect for an
infinitely extended cold nuclear medium (the term with $l_p$)
remains unchanged. One should remark, that the corrections due to
the finite survival probability of the prehadron are numerically
very small in general.
If one assumes big homogeneous nuclei for which $R_A\gg l_p$, the
finite size effects and the survival probability become negligible
and the acquired $\Delta p_\perp^2$ is given by
\begin{equation}\label{lin}
\langle\Delta p_\perp^2\rangle_q=\langle\sigma q_{\bot}^2\rangle\, l_p \, \rho_0.
\end{equation}

A hadron with momentum fraction $z_h$ has a $\langle\Delta
p_\perp^2\rangle_h$ reduced by $z_h^2$ compared to the quark
$\langle\Delta p_\perp^2\rangle_q$. This is a purely kinematical factor and
accounts for the fact that the average $\langle\Delta p_\perp\rangle_q$ is
shared among the produced hadrons according to their energy fractions (c.f. also Ref.~\cite{Kopeliovich:2006xy}).

\begin{equation}\label{pt_had}
\langle\Delta p_\perp^2\rangle_h= z_h^2 \langle\Delta p_\perp^2\rangle_q.
\end{equation}

\begin{figure}[h]
\includegraphics[width=0.5\textwidth]{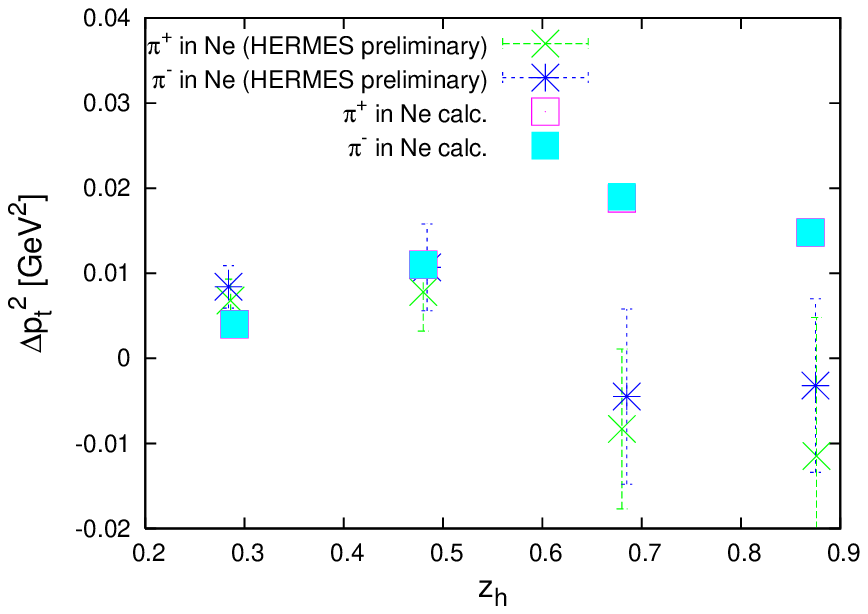}
\includegraphics[width=0.5\textwidth]{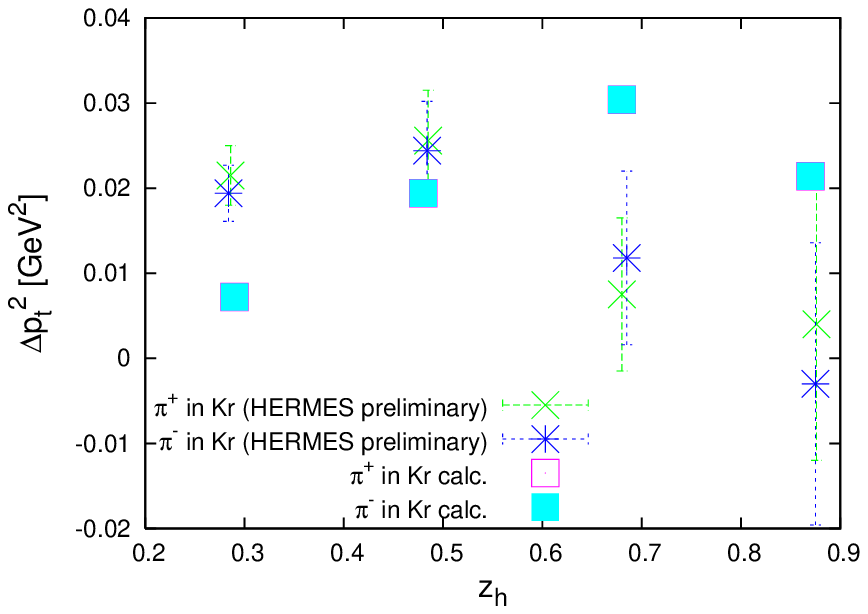}
\includegraphics[width=0.5\textwidth]{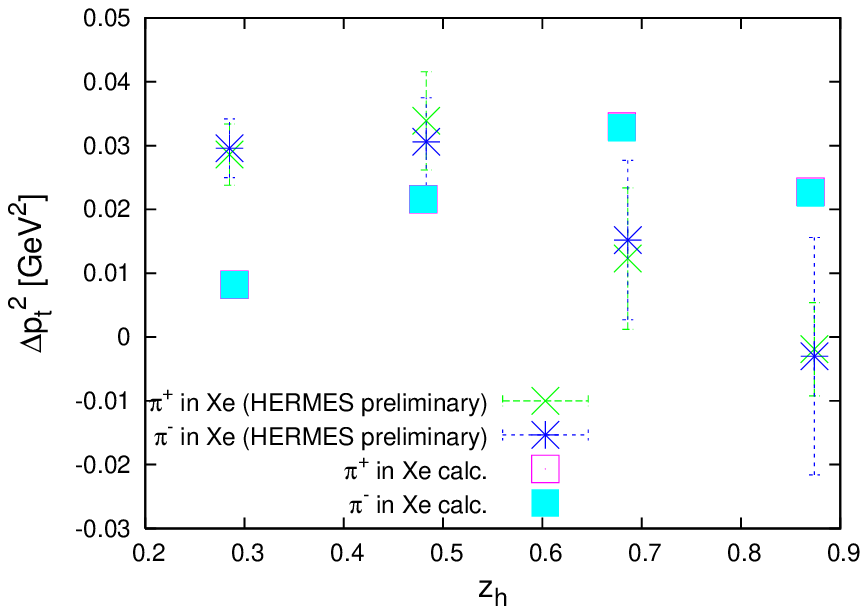}
\caption{\label{zh} $p_\perp$-broadening as a function of $z_h$ for pions in Ne, Kr and Xe.}
\end{figure}



For a more realistic computation of the transversal momentum
broadening of the quark, we use a Woods-Saxon distribution for the
nuclear density in the following. In Fig. \ref{zh}, we compare our
results for $p_{\bot}$ broadening with the HERMES preliminary data
\cite{VanHaarlem:2007kj} for $\pi^+$ and $\pi^-$ for the three
nuclei $^{20}$Ne, $^{84}$Kr and $^{132}$Xe as a function of $z_h$.
Since the prehadron formation length entering the calculation is a
function of $z_h$ and $\nu$, we take for the value of $\nu$ the
experimental average in the given $z_h$-bin. As one can see in the
plots for the dependence on $z_h$ there is qualitative agreement
between the calculation and the preliminary experimental data. The general
shape of the $z_h$-dependence has deficiences: For intermediate
$z_h$ the agreement is good in all three cases, but  in the small
$z_h$ bin the theoretical $\Delta p_\perp^2$ for Kr and Xe is too
small and  in the large $z_h$ bins it is too large. Furthermore,
our model does not differentiate between $\pi^+$ and $\pi^-$. In
order to do so, one would need to employ a more sophisticated
model which allows for flavor dependent prehadron formation
lengths (see Ref.~\cite{Accardi:2005jd}). Within the error bars,
the preliminary data do not discriminate  between $\pi^+$ and $\pi^-$ and the
expected effect seems to be small.

\begin{figure}[h]
\includegraphics[width=0.5\textwidth]{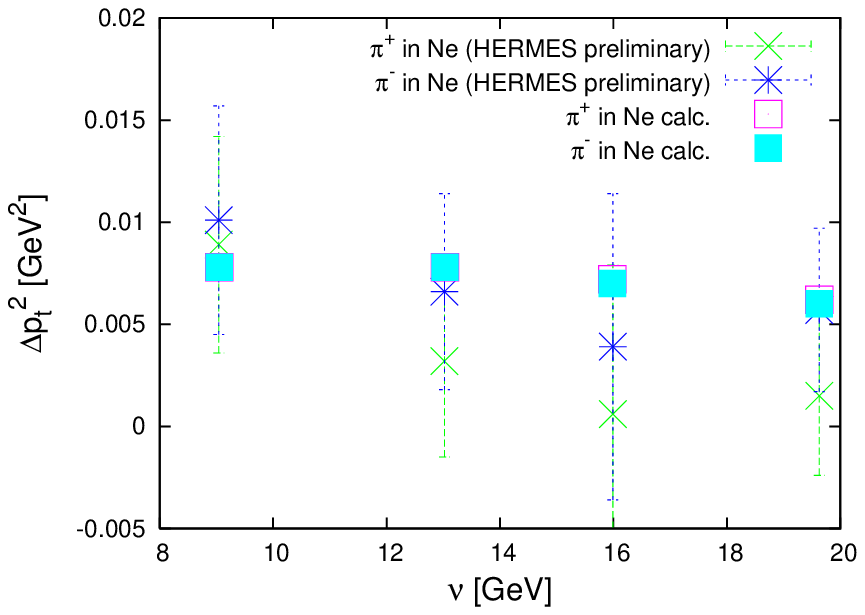}
\includegraphics[width=0.5\textwidth]{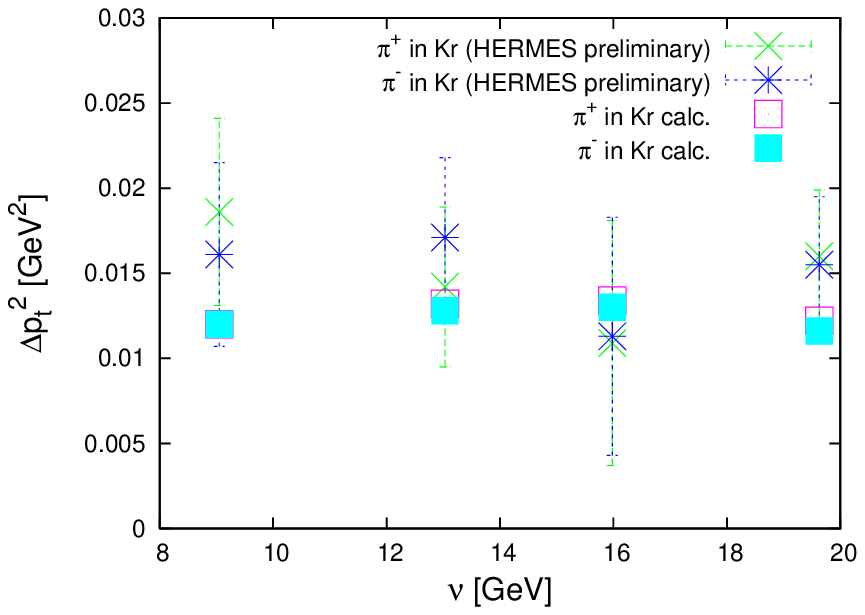}
\includegraphics[width=0.5\textwidth]{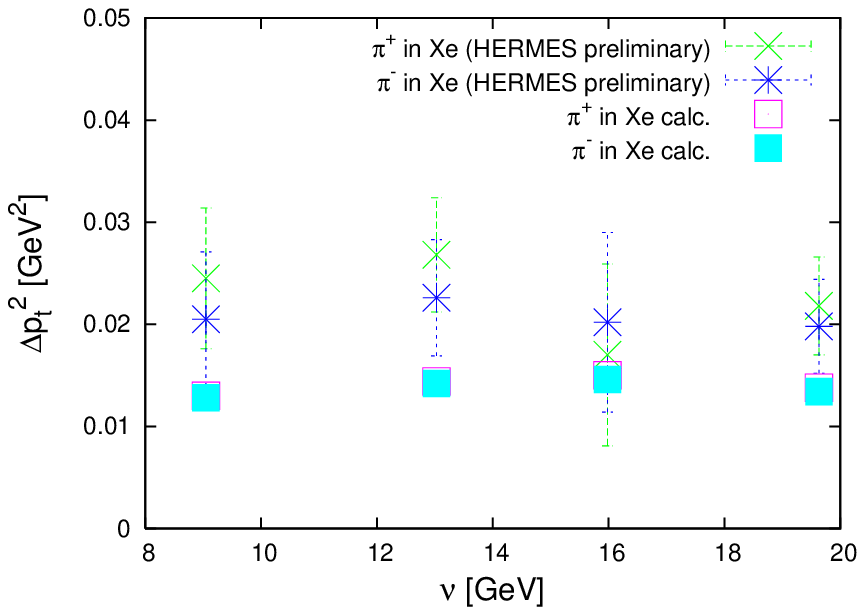}
 \caption{\label{nu} $p_{\bot}$-broadening of pions in Ne, Kr and Xe as a function of the photon energy $\nu$.}
\end{figure}

In Fig. \ref{nu} we display the
dependence of $\Delta p_{\bot}^2$ on the photon energy $\nu$.
Similar to the $z_h$ plot, we use in the computation of the prehadron formation length
the experimental mean value of $z_h$ in the given $\nu$-bin.
Because of the increase of formation time with $\nu$ one  would expect
that the $p_\perp$-broadening increases with photon energy.
In the preliminary data and in the
calculation, however,
the broadening $\Delta p_\perp^2$
decreases with $\nu$. We think that this is due to the experimental constraints on the kinematics.
With increasing
$\nu$ the experimental $\langle z_h \rangle$ (typically $z_h\simeq 0.35 - 0.45$ here) decreases
which lowers the resulting  $\Delta p_\perp^2$.
We remark that $\Delta p_\perp^2$ at constant $z_h$ increases  with $\nu$
in the preliminary CLAS data at J-Lab \cite {Brooks}. The much lower energy of DIS in this experiment,
however, may also allow hadron elastic scattering as an important source of nuclear
broadening.

\section{Hadronic $p_\perp$-broadening as  function of $Q^2$}

The variation of $p_{\bot}$- broadening with the photon virtuality
is the third piece of information we have from the preliminary data.
Recently an evolution equation has been constructed which
includes not only the splitting terms in the evolution of the
parton distributions but also a scattering term for the interactions
of the parton with the hot medium, i.e. the quark-gluon plasma \cite{Domdey:2008gp}.
We will use the equivalent equation in cold nuclear matter.
Soft collisions with the nucleons do not change the virtuality of the parton,
only increase its transverse momentum.
The ``fragmentation functions'' $ D_q^h(z_h,Q^2,\vec p_\perp)$ give the probability for an initial
quark  $q$ to convert into hadron $h$ with momentum fraction $z_h$,
virtuality $Q^2$ and transverse
momentum $\vec p_\perp$ in the course of the cascade.
The  evolution equation for this multiple differential function has the following form in vacuum \cite{Bassetto:1979nt}:
\begin{eqnarray}\label{bassetto}
\frac{ \partial D_q^h(z_h,Q^2,\vec p_\perp)}{\partial \log(Q^2)}
&=&\frac{\alpha_s(Q^2)}{2 \pi}
\int_{z_h}^1
\frac{dy}{y} P_q^r(y,\alpha_s(Q^2))\times\\ &&\hspace{-2cm}\int\frac{ d^2\vec q_\perp}{\pi}
\delta\left(y(1-y)Q^2-\frac{Q_0^2}
{4}-q_\perp^2\right)
D_r^h\left(\frac{z_h}{y},Q^2,\vec p_\perp-\frac{z_h}{y} \vec q_\perp\right)\nonumber
\end{eqnarray}
The above equation takes care of the mass constraint $y(1-y)Q^2=
Q_0^2/4+q_\perp^2$ arising in the splitting with momentum
fractions $y$ and $1-y$.  The transverse momenta appear together
with longitudinal momentum fractions to guarantee boost
invariance. After integration over transverse momentum one obtains
the standard DGLAP equation.  For electron-nucleus collisions we
consider the shower inside nuclear matter with a homogeneous nuclear density $\rho_0$.
We assume that the medium nucleons change the transverse momentum
of the quark by giving $\vec q_\perp$ kicks, but they do not
change its mass scale or virtuality.  Strictly speaking, this is only true
for small momentum transfers i.e. small angle scattering.  In
nuclear matter radiation is interleaved with scattering. Therefore
a scattering term $S(z_h,Q^2,\vec{p}_\perp)$ has to be added on
the right-hand side of Eq.~(\ref{bassetto}).  It has two parts, a
gain term for scattering into the given $z_h$-bin under
consideration and a loss term.

\begin{eqnarray}\label{Sder_full}
\lefteqn{
S(z_h, Q^2,\vec p_\perp)= \frac{\nu}{Q^2} \rho_0 \int _{z_h}^1 dw \int d^2\vec q_\perp
\frac{d\sigma}{d^2\vec q_\perp}}&&\nonumber\\
&&\times \left(D_q^h(w,Q^2,\vec p_\perp-w \vec q_\perp)-D_q^h(z_h,Q^2,\vec p_\perp)\right)
\delta\left(w- z_h-\frac{q_\perp^2}{2 m_b \nu}\right)
.
\end{eqnarray}

The evolution  allows to calculate two different  $\langle
p_\perp^2\rangle$ from the respective fragmentation functions in
the nucleus and in the vacuum Eq.~(\ref{bassetto})
\begin{equation}
\langle p_\perp^2\rangle =\frac{\displaystyle\int d^2 p_\perp
p_\perp^2 D(z,Q^2, p_\perp)}
    {\displaystyle\int d^2 p_\perp  D(z,Q^2, p_\perp)}.
\end{equation}
However, the $\langle p_\perp^2 \rangle$ defined in this equation
is solely coming from the evolution. The medium modification 
of the DGLAP evolution gives the difference of mean transverse momentum generated in
the evolution in the nucleus and in the vacuum. This piece is an
additional contribution to the multiple scattering contribution
$(\Delta p_{\bot}^2)_h(\bar Q^2)$  which we fix at $\bar Q^2=2.5$
GeV$^2$ to the preliminary data. In the calculation of the mean transverse
momentum broadening of the hadron the same averaged $\langle\sigma
q_{\bot}^2\rangle$ of the quark and the factor $z_h^2$ converting
quark to hadron transversal momentum squared appears naturally
together with transverse momentum integrated fragmentation
function of this specific hadron divided by this hadron
multiplicity.

\begin{equation}\label{ptofQ2}
(\Delta p_{\bot}^2)_h(Q^2)= (\Delta p_{\bot}^2)_h(\bar Q^2)+ z_h^2
\,\nu \,\rho_0 \, \langle\sigma q_\perp^2\rangle
\left(\frac{1}{\bar Q^2}-\frac{1}{Q^2}\right)\,.
\end{equation}
%
%

\begin{figure}[h]
\includegraphics[width=0.5\textwidth]{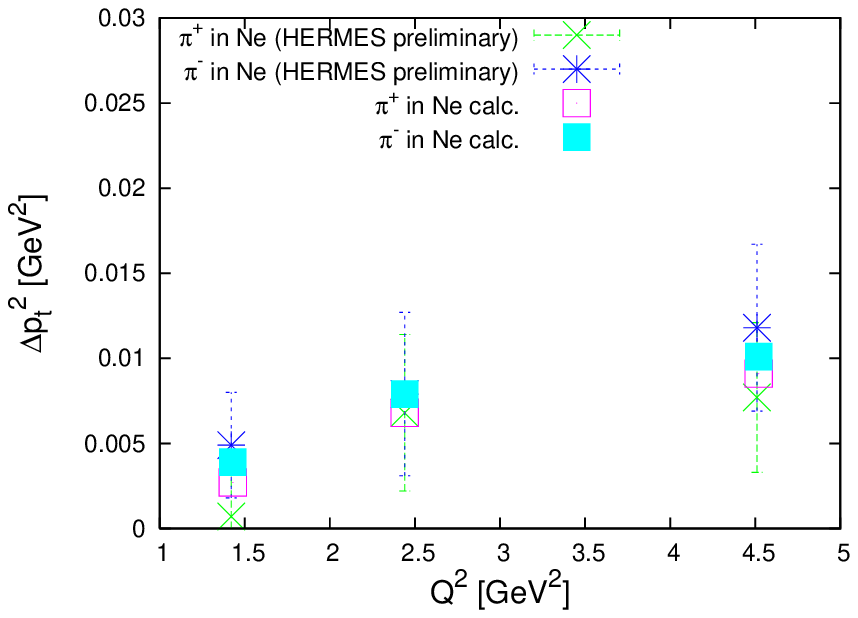}
\includegraphics[width=0.5\textwidth]{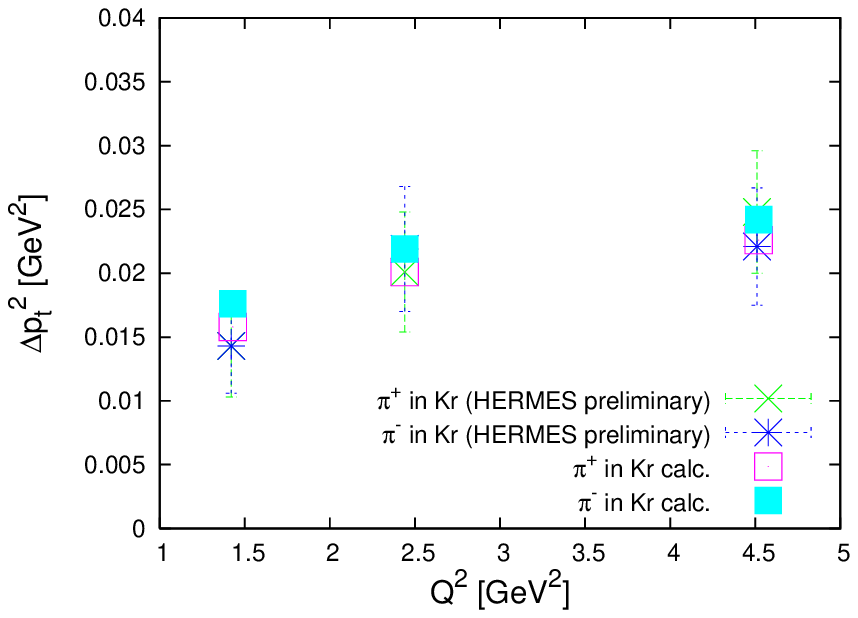}
\includegraphics[width=0.5\textwidth]{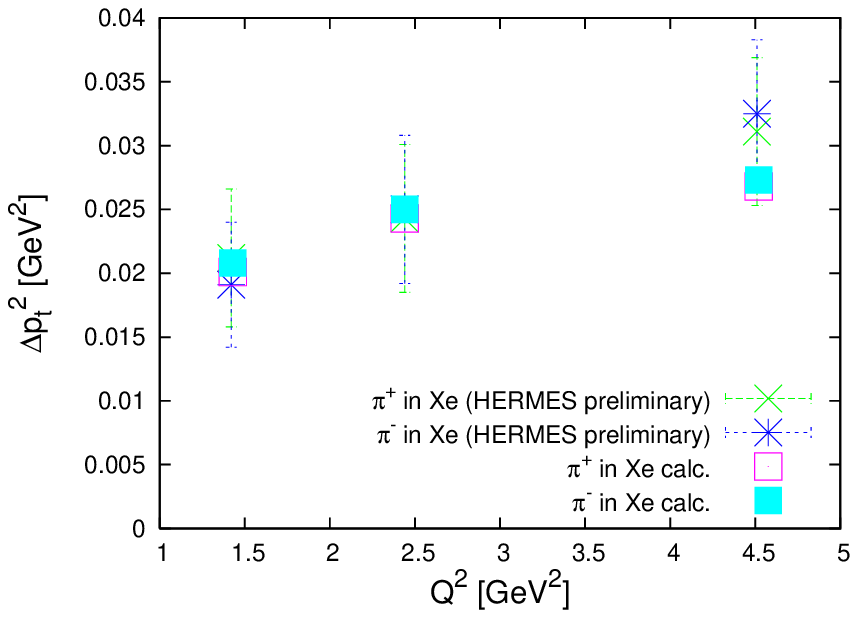}
 \caption{\label{Q2} $p_{\bot}$-broadening of pions in Ne, Kr and Xe as a function of the photon virtuality $Q^2$.}
\end{figure}

To lowest order, $\Delta p_\perp^2$ is
generated by the scattering term which gives a higher twist
contribution to the evolution from $\bar Q^2$ to $Q^2$. For
$Q^2>\bar Q^2$, the evolution enhances the mean $p_\perp^2$ and 
for $Q^2<\bar Q^2$ the devolution  decreases the mean $p_\perp^2$. 
Although HERMES preliminary data are for relatively small photon virtualities
of $Q^2=1.5-4.5$ GeV$^2$, this yields a sizeable effect for
$\Delta p_\perp^2$. 
As one sees in Fig. \ref{Q2} the calculated $Q^2$-dependence is in
good qualitative agreement with the preliminary data.  It is very encouraging
to see that the formalism of modified evolution equations which we
proposed for the quark-gluon plasma can also be related to deep
inelastic scattering on nuclei.

In the plasma, the corresponding scattering term was causing a
suppression of the fragmentation function in the medium i.e. jet
quenching \cite{Domdey:2008gp}. However, the situation in cold
nuclear matter is new: The scattering term is much smaller and the
medium fragmentation functions $D(z,Q^2)$ are almost unchanged.
To lowest order, the medium-induced $(\Delta p_{\bot}^2)_h(Q^2)$ is
originating only from the higher twist scattering term $\propto
1/Q^2$. Scaling violations due to higher twist effects have been
considered in \cite{Pirner:2006ui,Kopeliovich:2008td} for
processes with a large $z_h$.

The dependence of $\Delta p_\perp^2$ on $Q^2$
is also discussed in Ref. \cite{Accardi:2008fz}. This reference
also suggests two additional mechanisms as possible sources for
the $Q^2$-dependence, namely NLO processes like photon-gluon
fusion and possible colored prehadrons which lose energy by gluon
bremsstrahlung.

\section{Discussion}\label{sec:conclusions}

We have calculated $p_\perp$-broadening in transverse momentum
distributions from the dipole model and compared it to recent HERMES preliminary
data. We find qualitative agreement with the $z_h$-, $\nu$- and $Q^2$-dependences
of $p_\perp$-broadening. The dependence on the photon
virtuality has been calculated with a modified DGLAP evolution
equation. Finally we have estimated the effect of absorption of the
prehadronic state for $p_\perp$-broadening.

In a recent paper \cite{Accardi:2008fz} the nuclear multiplicity
ratio $R_M$ has been related to $p_\perp$ broadening, using a
similar picture of hadronization
\cite{Accardi:2002tv,Accardi:2005jd}. In this paper, the prehadron
formation time is extracted from the multiplicity ratio to
$t_p\equiv l_p\propto 0.8 \frac{\nu}{\kappa} z_h^{0.5} (1-z_h)$.
This formation time is similar to \eqnref{lp} but up to 30\%
smaller at mid $z_h$. On the other hand the difference between
hadronic and partonic $\Delta p_\perp^2$ is not spelled out.

There is another question which needs to be adressed: How do the
HERMES preliminary data \cite{VanHaarlem:2007kj} match with the preliminary CLAS
data \cite {Brooks}? The main difference between CLAS and HERMES is
the beam energy, which is $2-5$ GeV in CLAS in contrast to $7-23$ GeV
at HERMES. Therefore one expects that the prehadron formation time
$l_p$ is smaller by a factor $\gtrsim 3$ at CLAS and consequently also
the resulting hadronic broadening $\Delta p_\perp^2$ would be much
smaller. In the CLAS experiment, however, effects for hadronic $\Delta
p_\perp^2$ are of similar magnitude as in the HERMES experiment.  A
possible explanation can be that the prehadron stage contributes to
the hadronic broadening. At these low energies the pion-nucleon
elastic cross section is of the same magnitude as the inelastic cross
section. Therefore elastic scattering competes with absorption for the
outgoing prehadron. Since the angular distribution of pion-nucleon
scattering has still sizeable contributions from $u$-channel exchange,
large transverse momentum exchanges are possible. A good check is
possible when the whole angular distribution of the produced hadron is
measured.  Another important feature of the preliminary CLAS data is the linear
rise with $\nu$ which possibly saturates at $\nu\simeq 4$ GeV. This
linear rise is consistent with $\Delta p_\perp^2\propto l_p\propto
\nu$ as proposed in \eqnref{lin}.  As discussed, at CLAS hadronization may set in
inside the nucleus in contrast to HERMES. This does require also a
careful analysis  of the energy dependence of elastic
prehadronic scatterings. Therefore in our three stage model of
hadronization, the second and third step play a more important role.

\setcounter{equation}{0}

\vskip1cm

{\bf Acknowledgments}

This work was supported within the framework of the Excellence Initiative by the German
  Research Foundation (DFG)
  through the Heidelberg Graduate School of Fundamental Physics
  (grant number GSC 129/1), by Fondecyt (Chile) grant 1050589,
by DFG (Germany)  grant PI182/3-1 and by the Gesellschaft f\"ur Schwerionenforschung (GSI)
Darmstadt.

\vskip.15cm


\end{document}